\documentclass[letter]{aa}
\AddToHook{begindocument/before}{\usepackage{hyperref}}
\usepackage{graphicx}
\usepackage{txfonts}
\usepackage{fancyhdr}
\usepackage{xcolor}
\usepackage{lineno}
\usepackage{ulem}

\usepackage[title]{appendix}
\begin{document}

   \title{Evidence for a gamma-ray molecular target in the enigmatic PeVatron candidate \object{LHAASO J2108+5157}\thanks{As part of the thesis to be submitted by Toledano–Ju\'arez as a partial fulfillment for the requirements of Ph. D. Degree in Physics, Doctorado en Ciencias (F\'isica), CUCEI, Universidad de Guadalajara}
   }


   \author{E. de la Fuente,
          \inst{1,2\thanks{Sabbatical research stay at ICRR-UTokyo in 2021. Corresponding author }}
           I. Toledano--Ju\'arez,
          \inst{3}
           K. Kawata,
          \inst{2}
           M. A. Trinidad,
          \inst{4}
          M. Yamagishi,
          \inst{5}
          S. Takekawa,
          \inst{6}
          D. Tafoya,
          \inst{7}
          M. Ohnishi,
          \inst{2}
          A. Nishimura,
          \inst{8}
          S. Kato,
          \inst{2}
          T.Sako,
          \inst{2}
          M. Takita,
          \inst{2}
          \and
          H. Sano,
          \inst{9}
          \and
          R.K. Yadav
          \inst{10}
          }

   \institute{Departamento de F\'{i}sica, Centro Universitario de Ciencias Exactas e Ingenier\'{i}as, Universidad de Guadalajara, Blvd. Marcelino Garc\'{i}a Barrag\'an 1420, 44430, Guadalajara, Jalisco, M\'exico\\
              \email{eduardo.delafuente@academicos.udg.mx}
        \and      
              Institute for Cosmic Ray Research, University of Tokyo, Kashiwa 277-8582, Japan
        \and
              Doctorado en Ciencias en F\'{i}sica, Centro Universitario de Ciencias Exactas e Ingenier\'{i}as, Universidad de Guadalajara, Blvd. Marcelino Garc\'{i}a Barrag\'an 1420, 44430, Guadalajara, Jalisco, M\'exico
        \and
              Departamento de Astronom\'{i}a, Universidad de Guanajuato, Apartado Postal 144, 36000, Guanajuato, Guanajuato, M\'exico Department of Space, Earth, and Enviroment, Chalmers University of Technology, Onsala Space Observatory, 439 92 , Sweden 
        \and
              Institute of Astronomy, Graduate School of Science, The University of Tokyo, 2-21-1 Osawa, Mitaka, Tokyo 181-0015, Japan
         \and
             Department of Applied Physics, Faculty of Engineering, Kanagawa University, 3-27-1 Rokkakubashi, Kanagawa-ku, Yokohama, Kanagawa 221-8686, Japan  
         \and
             Department of Space, Earth, and Enviroment, Chalmers University of Technology, Onsala Space Observatory, 439 92 , Sweden
         \and
             Nobeyama Radio Observatory, National Astronomical Observatory of Japan (NAOJ), National Institutes of Natural Sciences (NINS), 462-2 Nobeyama, Minamimaki, Minamisaku, Nagano 384-1305, Japan
         \and
             Faculty of Engineering, Gifu University, 1-1 Yanagido, Gifu 501-1193, Japan
         \and    
             National Astronomical Research Institute of Thailand (Public Organization), 260 Moo 4, T. Donkaew, A. Maerim, Chiangmai, 50180, Thailand             }

\date{Received April 18, 2023; accepted June 22, 2023}

\nolinenumbers

  \abstract
   {Peta-eV (PeV) astronomy emerged in 2021 with the discovery of ultra-high-energy gamma-ray sources associated with powerful natural particle accelerators known as PeVatrons. In order to determine the nature of their emission, namely whether it has a hadronic or leptonic origin, it is essential to characterise the physical parameters of the environment where it originates.}
   {We unambiguously confirm the association of molecular gas with the PeVatron candidate \object{LHAASO J2108+5157} using unprecedented high angular-resolution (17$^{\prime \prime}$) $^{12,13}$CO($J$=1$\rightarrow$0) observations carried out with the Nobeyama 45m radio telescope.}
   {We characterised a molecular cloud in the vicinity of the PeVatron candidate \object{LHAASO J2108+5157} by determining its physical parameters from our $^{12,13}$CO($J$=1$\rightarrow$0) line observations. We used an updated estimation of the distance to the cloud, which provided a more reliable result. The molecular emission was compared with excess gamma-ray images obtained with \textit{FERMI}--LAT at energies above 2 GeV to search for spatial correlations and test a possible hadronic ($\pi^0$ decay) origin for the gamma-ray emission.}
   {We find that the morphology of the spatial distribution of the CO emission is strikingly similar to that of the \textit{FERMI}--LAT excess gamma ray. By combining our observations with archival 21cm HI line data, the nucleons (HI + H$_2$) number density of the target molecular cloud is found to be 133.0 $\pm$ 45.0 cm$^{-3}$, for the measured angular size of 0.55 $\pm$ 0.02$^\circ$ at a distance of 1.6 $\pm$ 0.1 kpc. The resulting total mass of the cloud is M(HI +H$_2$) = 7.5$\pm$2.9$\times$10$^3$ M$_{\odot}$. Under a hadronic scenario, we obtain a total energy of protons of W$_p$ = 4.3$\pm$1.5 $\times$ 10$^{46}$ erg with a cutoff of 700$\pm$300 TeV, which reproduces the sub-PeV gamma-ray emission.}
   {We identified a molecular cloud in the vicinity of LHAASO J2107+5157 as the main target where cosmic rays from an unknown PeVatron produce the observed gamma-ray emission via $\pi^0$ decay.}


\titlerunning{The molecular gamma-ray target of \object{LHAASO J2108+5157}}
\authorrunning{de la Fuente et al.}

\maketitle


\section{Introduction}

PeVatrons were recently discovered thanks to the highly sensitive gamma-ray observatories LHAASO--KM2A \citep[e.g.][]{Cao2021a}, Tibet-$\gamma$AS \citep[e.g.][]{Amenomori2021a,Amenomori2021b,Amenomori2021c}, and HAWC \citep[e.g.][]{Abeysekara2023,Abeysekara2021,Abeysekara2020}. The study of PeVatrons is important to understand the origin of the most energetic particles in the universe. As an emerging field in astrophysics there are two important issues to be addressed. The first is to determine the nature of the observed gamma-ray emission, leptonic (e.g. electrons in the inverse Compton effect) or hadronic (e.g. neutral pion decay through the interaction of cosmic-rays with molecular gas). The second is to identify the astronomical counterpart of the PeVatron, an object that accelerates the particles (at PeV) and then produces the observed gamma-ray emission in a target (e.g. molecular gas). Although several PeVatrons have already been associated with a counterpart \citep[e.g. \object{Boomerang}, \object{Cygnus Cocoon};][]{Amenomori2021a,Abeysekara2021}, there are exceptions, such as \object{LHAASO J2108+5157} (J2108 hereafter), whose counterpart remains elusive \citep{Cao2021b}. 

J2108 was proposed by \citet{Cao2021a,Cao2021b} as a PeVatron candidate with gamma-ray emission from 25 to 100 TeV (9.5 $\sigma$) and above 100 TeV (8.5 $\sigma$) measured with the LHAASO-KM2A observatory at an angular resolution of $\sim$0.3$^{\circ}$ in the \object{Cygnus OB7} molecular cloud (hereafter Cyg-OB7; \citealt{Reipurth2008} and references therein). It is considered one of the most enigmatic PeVatrons as it lacks a leptonic gamma-ray emitter and has a clear association with a molecular cloud (e.g. \object{[MML2017]4607}; \citealt{Cao2021b}, and \object{[FKT-MC]2022}; \citealt{delaFuente2023}; hereafter Paper I). Pioneering studies to better understand this enigmatic object have been carried out by \citet{DeSakar2023}; Paper I; \citet{Abe2022,Kar2022}; and \citet{Cao2021b}. Particularly, \citet{Abe2022} used the Fermi LAT 12-Year Point Source Catalog (4FGL-DR3) to analyse a region of interest around the J2108 position, which includes its high-energy (HE) counterpart \object{4FGL J2108.0+5155}. After removing \object{4FGL J2108.0+5155} from the source model and using a putative source with a power-law spectrum, they found a new source exhibiting a hard photon index. This source was called HS for hard source.

\begin{figure*}[!ht]
\begin{center}

\includegraphics[width=\textwidth]{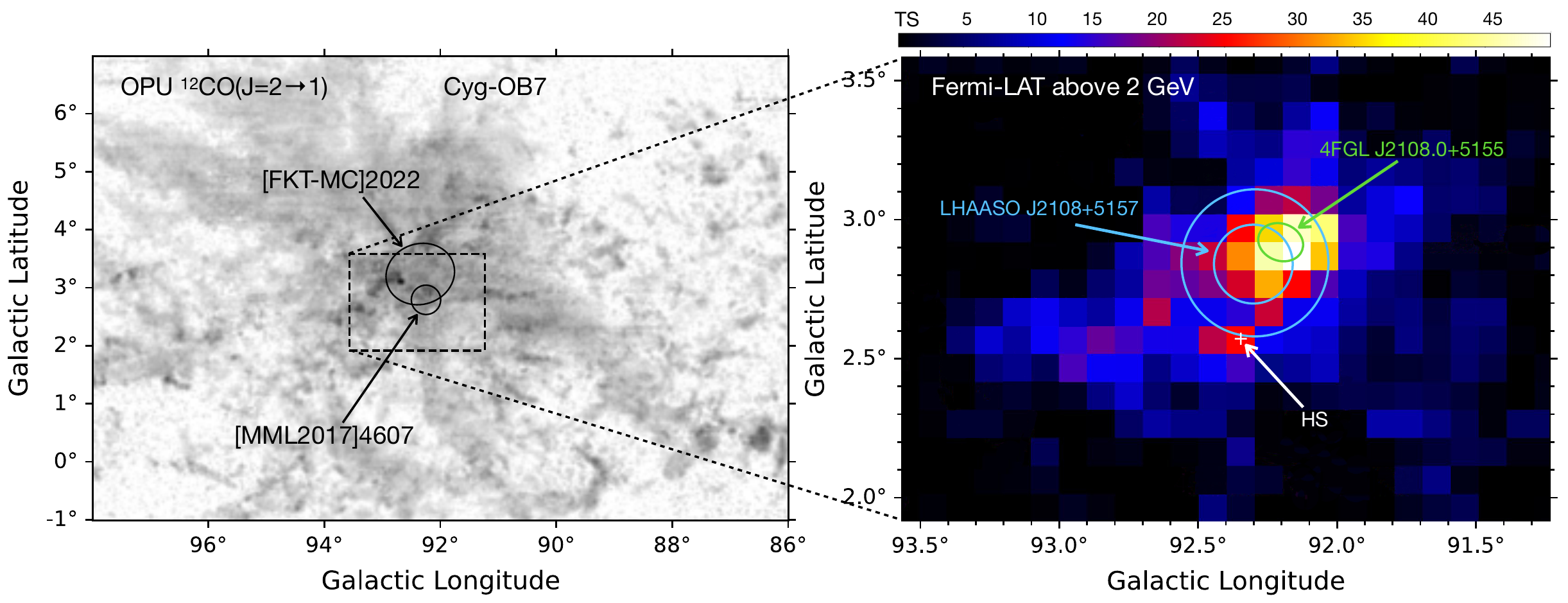}
\end{center}

\caption{Molecular clouds and gamma-ray sources associated with J2108. {\bf Left:} OPU $^{12}$CO($J$=2$\rightarrow$1) emission map towards the Cyg-OB7 association \citep{delaFuente2023}. The molecular clouds \object{[MML2017]4607} and \object{[FKT-MC]2022} are indicated with ellipses. The angular resolution of the map is 3$^{\prime}$. {\bf Right:} \textit{Fermi}-LAT TS map above 2 GeV in the region around J2108 adapted from Fig.~3 of \citet{Abe2022}. The 95\% positional uncertainty of 4FGL J2108.0+5155 is shown as a green ellipse. The cyan circles (radius of $0.14^\circ$ and $0.26^\circ$) correspond to the 95\% positional uncertainty and 95\% UL of J2108, respectively.}
\label{fig:Fermi_CygOB7}
\end{figure*}

In Paper I we analysed the molecular gas around J2108 using observations of $^{12,13}$CO($J$=2$\rightarrow$1) emission towards Cyg-OB7 at an angular resolution of $\sim$3$\arcmin$, taken with the Osaka Prefecture University (OPU) 1.85 m radio telescope \citep{Nishimura2020}. From these observations we find that in addition to the molecular cloud \object{[MML2017]4607}, there is another cloud, \object{[FKT-MC]2022}, that could be the target responsible for the emission detected with LHAASO. In the left panel of Figure~\ref{fig:Fermi_CygOB7} we show the OPU $^{12}$CO($J$=2$\rightarrow$1) moment 0 map of Cyg-OB7 indicating the locations of \object{[MML2017]4607} and \object{[FKT-MC]2022} (see Paper I). The right panel shows the test statistics (TS) map ($>$2 GeV) of the region studied by \citet{Abe2022} (see their Fig. 3). The 95\% positional error and 95\% upper limit (UL) on the source extension of J2108 (0.14$^{\circ}$ and 0.26$^{\circ}$, respectively) are shown as cyan circles. An ellipse indicating the 95\% positional error of 4FGL J2108.0+5155 and the location of the source HS are also shown in this panel.

The distance to J2108 was estimated by calculating the kinematic distance to the molecular clouds with which it is associated. The values for the distance range from 1.7 kpc to 3.3 kpc \cite[Paper I;][]{Cao2021b, Abe2022}. The difference in the values of the distances is partly due to outdated parameters used in the calculations of rotation curves, but also to ambiguity in the systemic velocity of the molecular clouds associated with J2108. In addition, the lack of consideration for measurements such as geometrical parallax to nearby sources adds to the uncertainty. As a result, the determination of physical parameters is accompanied by significant uncertainties. To obtain more reliable values for these parameters, updating the distance value to J2108 is imperative.

As a follow-up of Paper I, we present observations of $^{12,13}$CO ($J$=1$\rightarrow$0), hereafter $^{12,13}$CO, taken with the 45 m radio telescope at the Nobeyama Radio Observatory (NRO). The higher resolution of the present observations ($\sim$17$\arcsec$) enables us to better identify structures within the molecular cloud, which is not possible from previous surveys \citep[e.g.][]{Dame2001}. More importantly, the achieved sensitivity allows us to detect spectral components with the optically thin $^{13}$CO emission that were not feasible with previous observations \citep[Paper I;][]{Nishimura2020}. Using optically thin emission rather than being limited to optically thick emission permits a more reliable estimation of the CO column density used to calculate the physical parameters of the molecular cloud.

\section{Observations}
\label{sec:obs}

The $^{12,13}$CO observations were performed on February 1, 3, 4, 11, 12, and March 19--20, 2023 with the FOREST receiver \citep{Minamidani2016}. The mapped region has a size of $\sim$1 deg$^2$, and is centred at the galactic coordinates ($l,b$) = 92.307, 2.836, which corresponds to ($\alpha_{J2000},\delta_{J2000}$) = 21:08:52.92, 51:57:00.91. The local standard of rest (LSR) velocity covered the range $-$80<$V_{\rm LSR}$<80 km s$^{-1}$ with a spectral resolution of $\sim$0.5 km s$^{-1}$. The effective spatial resolution is 17$''$, and the main beam efficiency is 39$\pm$3\%.  The total integration time is 870 minutes covering 660 scans. Pointing errors were corrected to less than 5$''$ every 1.5 hours by observing a SiO maser source, IRAS21086+5238. 
Data reduction and calibration were performed according to the standard procedures using NOSTAR software provided by NRO\footnote{\url{https://www.nro.nao.ac.jp/~nro45mrt/html/obs/otf/export-e.html}} and described in \citet{Yamagishi2018}.  

We retrieved atomic hydrogen (HI) 21 cm line observations performed by the Dominion Radio Astrophysical Observatory\footnote{DRAO is part of the Canadian Galactic Plane Survey Project (CGPS). \url{https://www.cadc-ccda.hia-iha.nrc-cnrc.gc.ca/en/search/\#resultTableTab}} (DRAO; \citealt{Taylor2003}). The images are projected into a 1024$\times$1024 mosaic with a pixel size of 18$''$  and an angular resolution of 1$'$. The velocity resolution is 0.82 km s$^{-1}$, and the rms of the brightness temperature is between 2.1 and 3.2 K.

\section{Results and discussion}
\label{sec:results_discussion}

\begin{figure*}[!ht]
\begin{center}
\includegraphics[width=\textwidth]{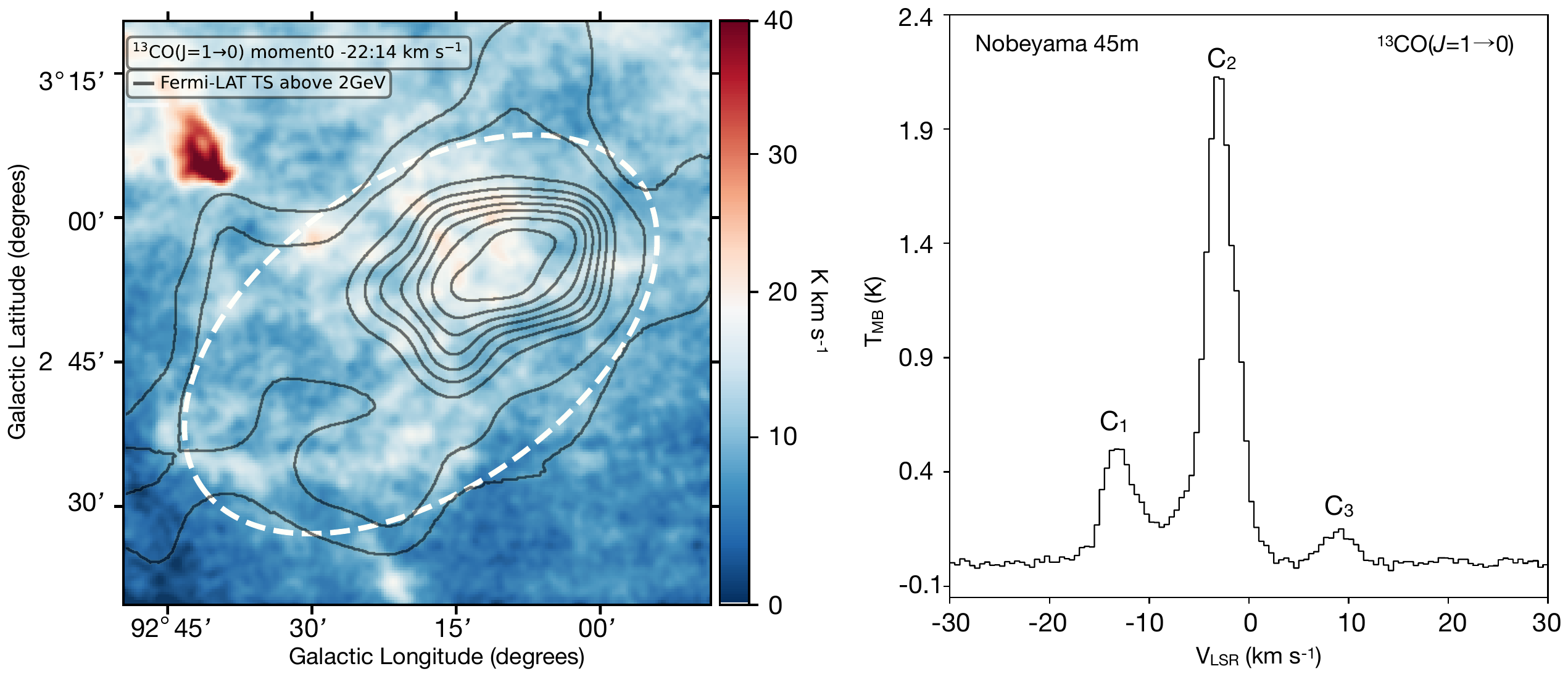}

\end{center}
\caption{{$^{13}$CO emission towards the vicinity of J2108.} \textbf{Left}: $^{13}$CO moment-0 map, integrated between $-$22 and 14 km s$^{-1}$, presented in a colour scale. The temperature scale has been corrected for main-beam efficiency. The \textit{Fermi}-LAT TS map from Fig~\ref{fig:Fermi_CygOB7} is overlaid with contours [10, 15, 30, 40, 50, 60, 70, 80, 90] in percentage of the maximum excess value. \textbf{Right}: Spectrum of $^{13}$CO ($J$=1$\to$0) emission extracted from the area outlined by the dashed ellipse in the left panel, which covers the Fermi-LAT gamma-ray region of excess. Three spectral components are identified: C$_1$, C$_2$ and C$_3$ centred at $\sim$ $-$13, $-$3, and $+$9 km s$^{-1}$ respectively.}
\label{fig:mom0-fermi-spectrum}
\end{figure*}

\begin{figure*}[!ht]
\begin{center}

\includegraphics[width=\textwidth]{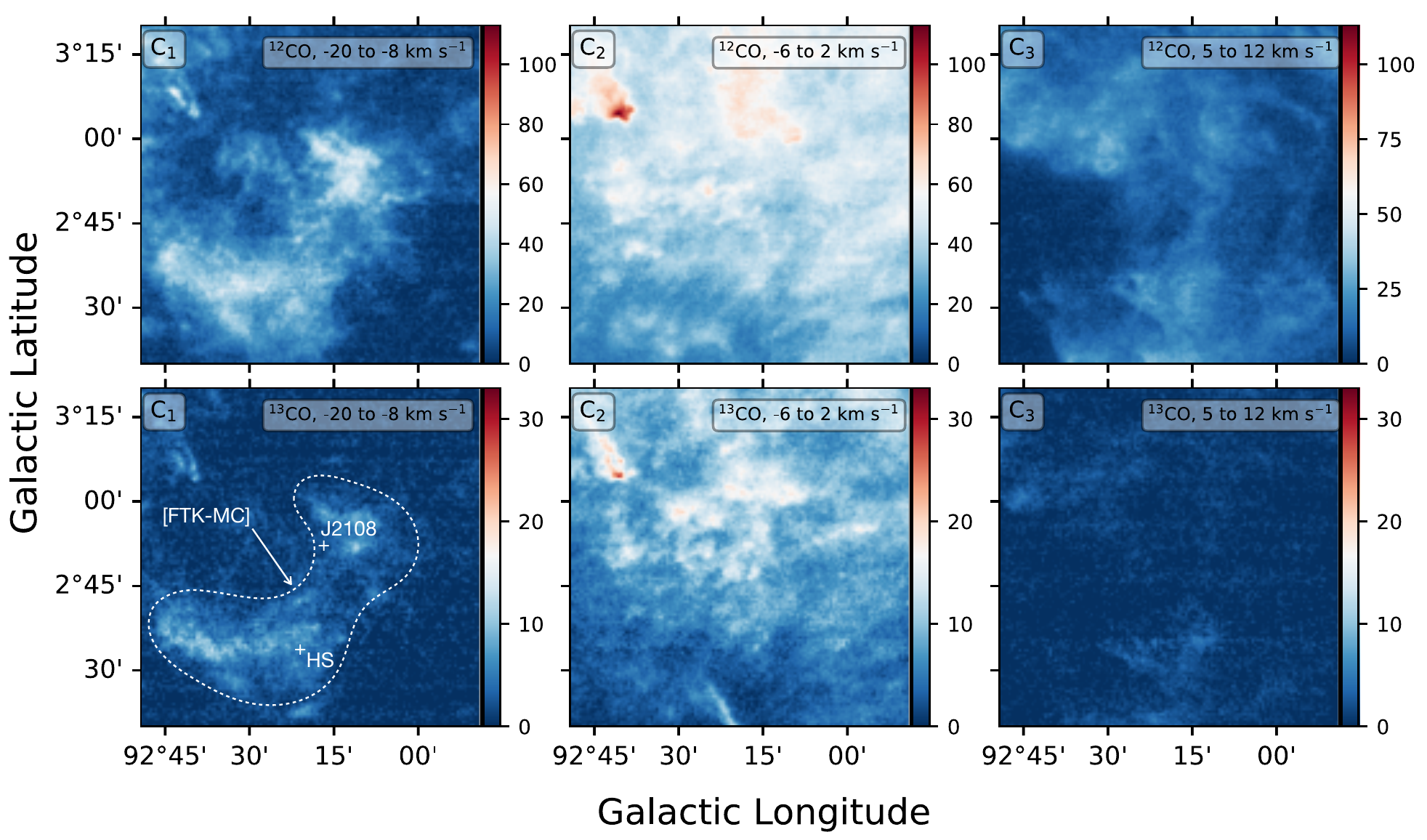}

\end{center}
\caption{
$^{12,13}$CO moment-0 maps of the three spectral components shown in {the right panel of} Fig.~\ref{fig:mom0-fermi-spectrum}. The units of the colour-scale are K~km~s$^{-1}$, corrected for antenna main-beam efficiency. The molecular cloud [FTK--MC] is most prominent in the map with velocity in the range from --20 to 8~km~s$^{-1}$, which corresponds to the spectral component C$_{1}$, and is delineated with a dashed line. The positions of the sources J2108 and HS are indicated with crosses.}  
\label{fig:spectra_map}
\end{figure*}

\begin{figure}[!ht]
\begin{center}

\includegraphics[width=\columnwidth]{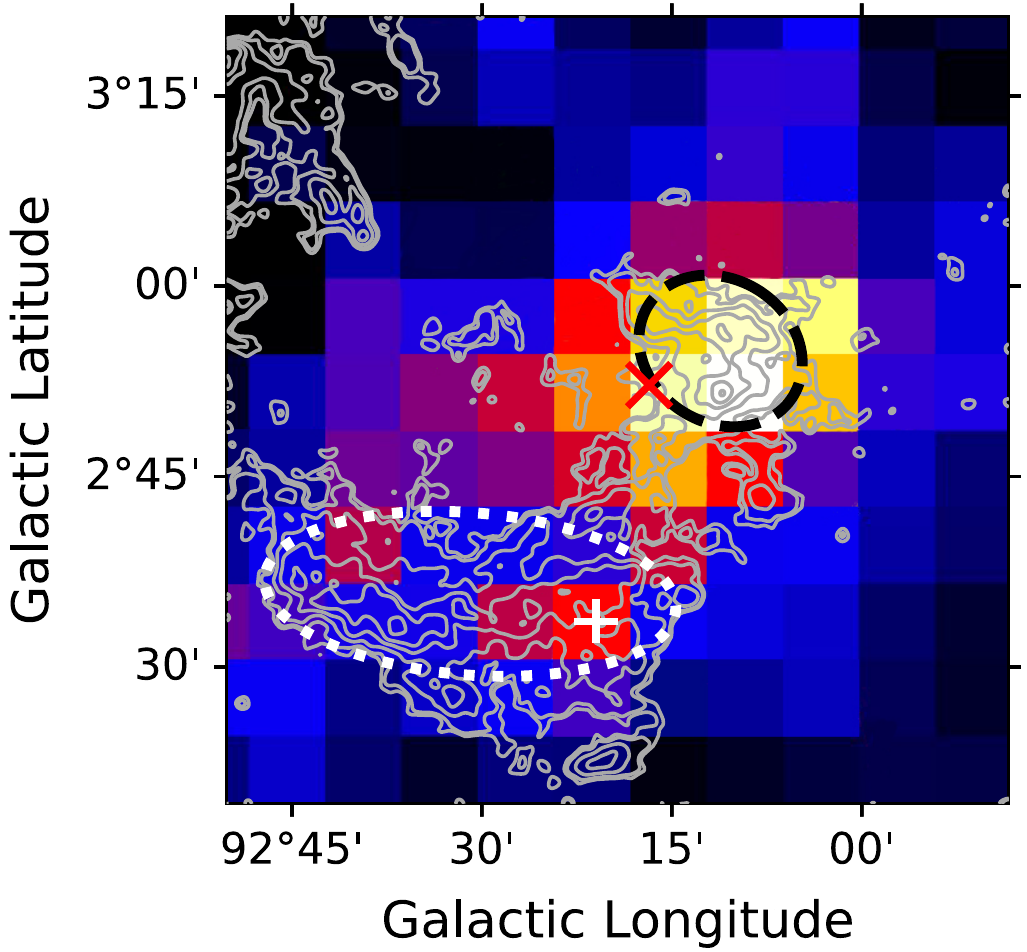}

\end{center}
\caption{\textit{Fermi}-LAT map (in colours) from Fig.~\ref{fig:Fermi_CygOB7} with$^{13}$CO moment 0 map overlaid as grey contours. The $^{13}$CO map was integrated between -20 and -8 km s$^{-1}$. The contours are [-4,4,5,8,12,16,20] times the rms value of 0.5 K km s$^{-1}$. The dashed black ([FTK--MC]J2108) and white dotted ([FTK--MC]HS) ellipses indicate the two regions where the physical parameters were determined. The croos (x) and the plus sign (+) cross indicate the central position of J2108 and HS, respectively. }
\label{fig:Fermi_Molecular}
\end{figure}

The left panel of Figure~\ref{fig:mom0-fermi-spectrum} shows the \textit{Fermi}-LAT TS map (contours) superimposed on the Nobeyama $^{13}$CO moment-0 map (colour map). From this map it can be seen that the spatial distribution of the CO emission exhibits a remarkable agreement with the morphology of the \textit{Fermi}-LAT gamma-ray excess. The CO emission is particularly bright\footnote{The brightest emission at the top left corner of the map corresponds to the star cluster region Kronberger 82.} at the locations of the peak of the \textit{Fermi}-LAT gamma-ray excess, and in the locations of J2108, 4FGL J2108.0+5155, and HS. The agreement of these morphologies indicates that the gamma-ray emission probably originates within the molecular cloud. Furthermore, if we assume that the LHAASO sub-PeV emission also traces the \textit{Fermi}-LAT gamma-ray excess morphology, this would mean that the gamma-ray emission results from the interaction between the gas of the molecular cloud and HE particles accelerated by a PeVatron.

The $^{13}$CO spectrum of the region encircled by the dotted ellipse is shown in the right panel of Fig.~\ref{fig:mom0-fermi-spectrum}. We identified three spectral components, labeled C$_1$, C$_2$, and C$_3$, with V$_{\rm LSR}$ $\sim$ $-$13, $-$3 and $+$9~km~s$^{-1}$, respectively. Figure~\ref{fig:spectra_map} shows the $^{12}$CO (top) and $^{13}$CO (down) moment-0 maps of these spectral components. These maps are integrated from $-$20 to $-$8 km s$^{-1}$ (left), from $-$6 to $+$2 km~s$^{-1}$ (middle), and from $+$5 to $+$12 km~s$^{-1}$ (right). The spatial correlation between the CO emission and the \textit{Fermi}-LAT gamma-ray excess is particularly evident for the molecular gas associated with the spectral component C$_1$ (see left panel of Fig.~\ref{fig:spectra_map}), hereafter referred to as [FTK--MC]. The cloud [FTK--MC] has an irregular morphology, consisting of two main components. Given their proximity to the gamma-ray sources J2108 and HS, we call these components [FTK--MC]J2108 and [FTK--MC]HS, located towards the north and south, respectively. The emission in the moment-0 maps of the central column of Fig.~\ref{fig:spectra_map} (C$_2$) covers the entire region (including \object{Kronberger 82}; Paper I). Although this emission partially arises from gas associated with \object{[FKT--MC]2022} (V$_{\rm LSR}$ $\sim$ --3 km s$^{-1}$; Paper I), it is difficult to quantify how much of it is due to \object{[FKT--MC]2022} and how much to ambient gas (see e.g. position--velocity diagrams in Figure~\ref{fig:1213CO_pv} Appendix \ref{appendix:spectra}). The emission corresponding to the spectral component C$_3$ seems to be associated with another molecular region coincident with the edge of an HI cloud observed with DRAO between 5 and 12 km s$^{-1}$ (see Appendix \ref{appendix:distance}). Given the spatial correlation between the emission of C$_1$ and the \textit{Fermi}-LAT gamma-ray excess, in this work we only consider the gas associated with this spectral component and exclude the contribution from the other two (see Appendix A). In Figure~\ref{fig:Fermi_Molecular} the $^{13}$CO emission (contours) of the cloud [FTK--MC] (i.e. the $^{13}$CO emission of the spectral component C$_{1}$) is superimposed on the \textit{Fermi}--LAT TS map. It is clear that the peak of the \textit{Fermi}--LAT gamma-ray excess coincides with the component [FTK--MC]J2108. 

All previous studies of the molecular gas around J2108 have used observations of the $^{12}$CO emission. The main problem with this approach is that, since the emission is optically thick, the derived column density is just a lower limit of the actual value, which hinders the determination of the density of nucleons. Our Nobeyama observations of the $^{13}$CO emission allow us to determine for the first time the physical parameters of the molecular gas around J2108 in the optically thin regime, resulting in more reliable values of the physical parameters. The calculation of the density of nucleons, n(H) = 2n(H$_2$) + n(HI), requires the size of the molecular cloud [FTK--MC]. However, given its complex morphology, we first fitted 2D Gaussian functions to the emission of [FTK--MC]J2108 and [FTK--MC]HS. The FWHM of the fitted Gaussians are represented as dashed and dotted ellipses in Fig.~\ref{fig:Fermi_Molecular}. The central positions of the fitted Gaussians are ($l,b$)=92.20$^\circ$, 2.90$^\circ$ and ($l,b$)=92.53$^\circ$, 2.59$^\circ$, respectively. A representative angular size for [FTK--MC] can be obtained as the sum of the fitted sizes of [FTK--MC]J2108 and [FTK--MC]HS, which gives a value of 0.55 ± 0.02 deg. Subsequently, the column densities of the molecular and atomic gas are obtained from the $^{13}$CO and HI emission corresponding to the spectral component C$_{1}$ (V$_{\rm LSR} \sim$ $-$13 km s$^{-1}$). The details of the data analysis and calculations are given in Appendix \ref{appendix:spectra}. 

For our analysis we adopted a distance of 1.6 $\pm$ 0.1 kpc. This distance was determined using the using the Revised Kinematic Distance Calculator of \citet{Reid2014,Reid2019} on the basis of the systemic velocity of the spectral component C$_{1}$ (V$_{\rm LSR} \sim$ $-$13 km s$^{-1}$). This value is similar, albeit slightly lower, to that adopted in Paper I (1.7~kpc), and close to the upper limit of the distances that \cite{Schneider2006} report for \object{Cygnus-X} . On the other hand, it is just half the value of 3.28 kpc proposed by \citet{Cao2021b}. The reason for adopting a distance of 1.6 $\pm$ 0.1 kpc is that the calculator takes into account the likelihood of the cloud being associated with nearby sources whose distance has been accurately measured via trigonometric parallax. In addition, the distance calculator of \citet{Reid2014,Reid2019} also takes into account the probable association of the molecular cloud with Galactic spiral arms, which increases the reliability of the estimated distance. The details of the calculation of the distance are given in Appendix \ref{appendix:distance}. Following the methods and equations presented in Paper I, and using the observational values associated with component C$_1$ shown in Table~\ref{table:Column1}, we derived physical parameters for [FTK--MC]J2108 and [FTK--MC]HS, separately. The column and volumetric densities of the cloud [FTK--MC] are taken as the average of the values of those obtained for [FTK--MC]J2108 and [FTK--MC]HS. All the derived physical parameters are listed in Table~\ref{table:Column2}.

We derived a nucleon number density of n(H) = 133 cm$^{-3}$, which four times higher than the value obtained by \cite{Cao2021b} of n(H) $\sim$ 30 cm$^{-2}$. After re-scaling our calculations with their proposed distance of 3.28 kpc, the number density obtained in this work is still a factor of 2 higher. This shows the importance of using optically thin emission to derive the physical parameters of the molecular gas. By considering the neutral pion decay hadronic model in the Naima software package \footnote{\url{https://naima.readthedocs.io/en/latest/index.html}} \citep[][and references therein]{Zabalza2015}, and using the estimated nucleon number density and distance as input parameters, and the additional parameters described in Paper I, we obtained a total required energy of the cosmic-ray proton population of W$_p$ $\sim$ 1.1 $\times 10^{47}$ ergs to reproduce the observed J2108 sub-PeV energy flux \citep{Cao2021b}. The parameters and results of the hadronic modelling are presented in Table \ref{tab:hadron}. 

We note that although the total energy of cosmic ray protons derived from our calculations is a factor of 20 lower than that obtained by \cite{Cao2021b}, it is consistent with the proposed scenario that PeVatron may be associated with an old supernova-like explosion \citep{Kar2022}. The energy of protons required to reproduce the LHAASO gamma-ray emission could easily be created by such a mechanism. However, we do not rule out the possibility of alternative astrophysical objects responsible for the acceleration of the HE protons. Further observations and studies will be of great importance to identify the nature of the PeVatron in this enigmatic source.


\section{Conclusions}
\label{sec:conclusion}

In the following we summarise the results and conclusions that we obtained from our Nobeyama $^{12,13}$CO($J$=1$\rightarrow$0) observations towards the region around sub-PeV gamma-ray source J2108: 

\begin{enumerate}

\item We identified for the first time a molecular cloud, [FTK--MC], whose location coincides with the position of the PeVatron candidate \object{LHAASO J2108+5157}. The morphology of this cloud is in striking agreement with the distribution of the \textit{Fermi}-LAT up to 2 GeV gamma-ray excess. 
      
\item  The cloud [FTK--MC] consists of two main components, [FTK--MC]J2108 and [FTK--MC]HS. The systemic velocity of [FTK--MC] is $\sim$ $-$13 km s$^{-1}$. It has an angular size of $\sim$0.55$^{\circ}$ and is located at a distance of 1.6 $\pm$ 0.1 kpc. 

\item The nucleon density, n(H) = 2n(H$_2$) + n(HI), in [FTK--MC], derived from optically thin $^{13}$CO emission, is estimated to be 133 cm$^{-3}$, which results in a total mass M(HI +H$_2$) $\sim$7.5$\times$10$^3$ M$_{\odot}$. The required total energy of protons to produce the observed sub-PeV emission of \object{LHAASO J2108+5157} is W$_p$ of 1.1 $\times$ 10$^{47}$ erg.

\item Based on these results, we favour a scenario where the molecular cloud [FTK--MC] is the main target of HE particles accelerated by an unidentified PeVatron. Thus, the gamma-rays observed by Fermi-LAT and LHAASO--K2MA have a hadronic component in nature.

\end{enumerate}

\begin{acknowledgements}

This work is based on observations with the 45 m telescope at the Nobeyama Radio Observatory (NRO). NRO is a branch of the National Astronomical Observatory of Japan, National Institutes of Natural Sciences. This research was supported by the Inter-University Research Programme of the Institute for Cosmic Ray Research, University of Tokyo, grant 2023i--F--005. IT--J acknowledges support from Consejo Nacional de Ciencias y Tecnolog\'ia, M\'exico (CONACyT) grant 754851, and Onsala Space Observatory during an academic stay in 2023. We are grateful for the computational resources and technical support provided by the Centro de An\'alisis de Datos y Superc\'omputo  (CADS) of the Universidad de Guadalajara through the Leo-Atrox supercomputer.

\end{acknowledgements}

%
%

\begin{appendix}

\section{Analysis of the spectra of the molecular emission}
\label{appendix:spectra}

\begin{figure}[!hb]
\begin{center}

\includegraphics[width=0.7\columnwidth]{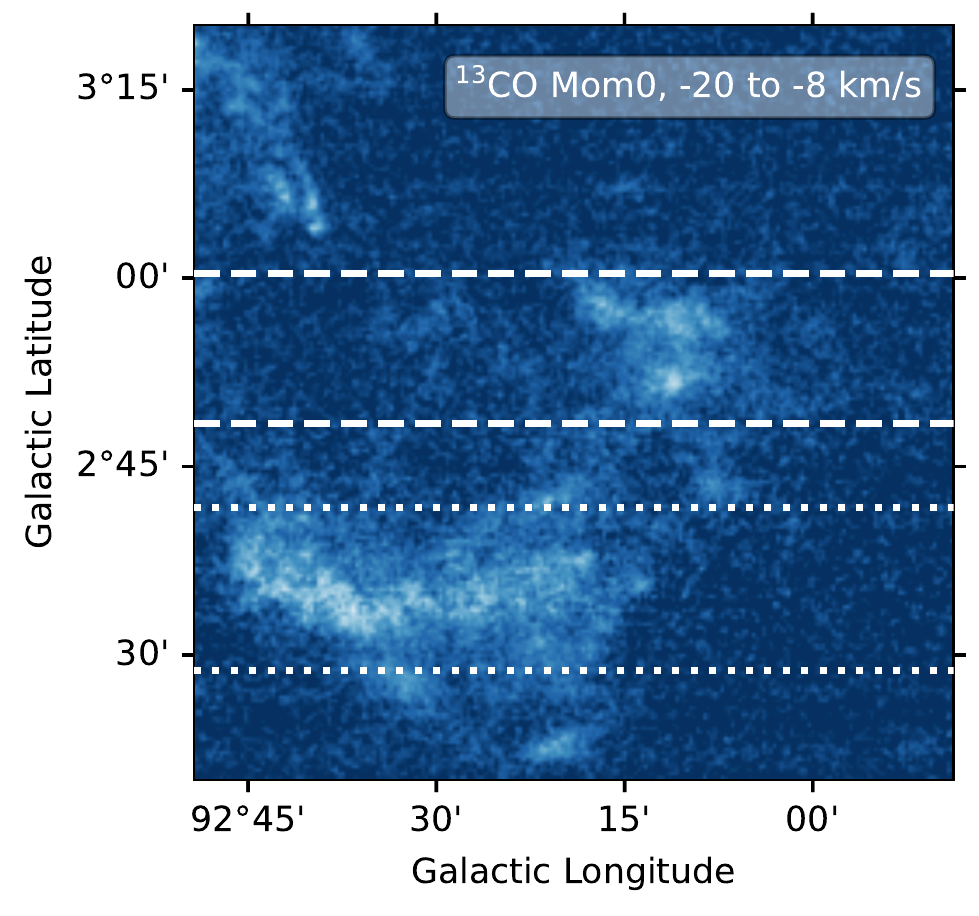}
\includegraphics[width=\columnwidth]{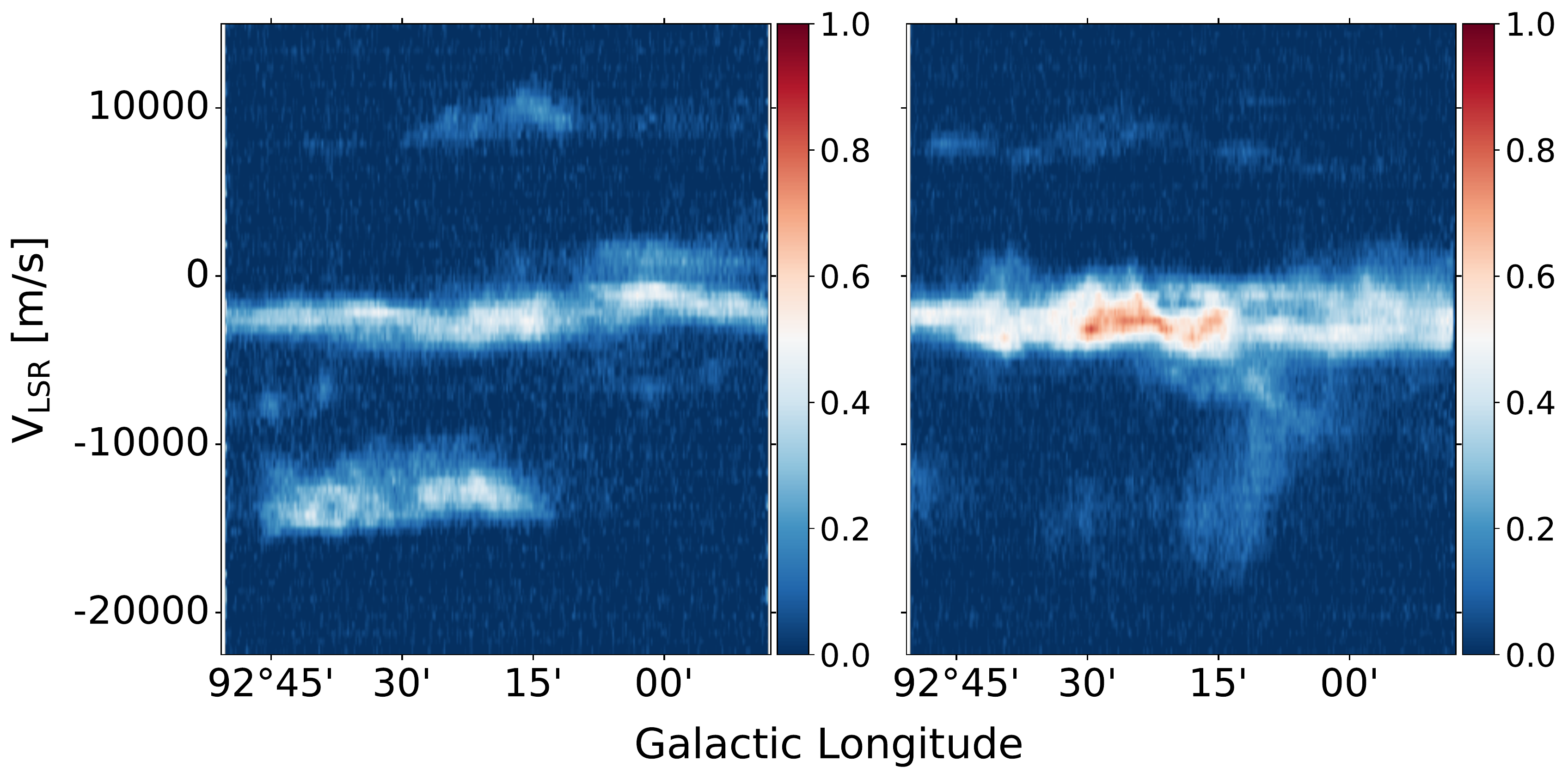}

\end{center}

\caption{$^{13}$CO map from Fig.~\ref{fig:spectra_map} top panel. The dashed and dotted rectangles indicate the ranges in the J2108-- and HS-- cloud positions respectively, where the galactic longitude as function of the V$_{\rm LSR}$ map is obtained: HS--cloud (bottom left) and J2108--cloud (bottom right). Three gas patches (C$_1$ bottom, C$_2$ middle and C$_3$ top) are observed. Colour--bar units are K deg.}
\label{fig:1213CO_pv}
\end{figure}

The lower panels of Fig.~\ref{fig:1213CO_pv} show two maps of Galactic longitude as a function of V$_{\rm LSR}$, position--velocity diagrams, by integrating emission within the Galactic longitude range that corresponds to the extent of the fitted ellipses of [FTK--MC]HS (dotted lines) and the [FTK--MC]J2108 (dashed lines), respectively (see Fig.~\ref{fig:Fermi_Molecular}). The molecular gas associated with each of the three spectral components C$_1$, C$_2$, and C$_3$  appears as horizontal bands in these PV diagrams. [FTK--MC]J2108 and [FTK--MC]HS are associated with molecular gas at V$_{\rm LSR}$ $\sim$ --13 km, corresponding to spectral component C$_1$, with [FTK--MC]J2108 (bottom right) showing a higher velocity dispersion of this component. Nevertheless, the spectral component C$_2$ covers the entire map at V$_{\rm LSR}$ $\sim$ --3 km s$^{-1}$, which can also be seen in Fig.~\ref{fig:spectra_map}. For this reason, we neglect component C$_2$ in the analysis.

\begin{figure}[!hb]
\begin{center}

\includegraphics[width=0.49\columnwidth]{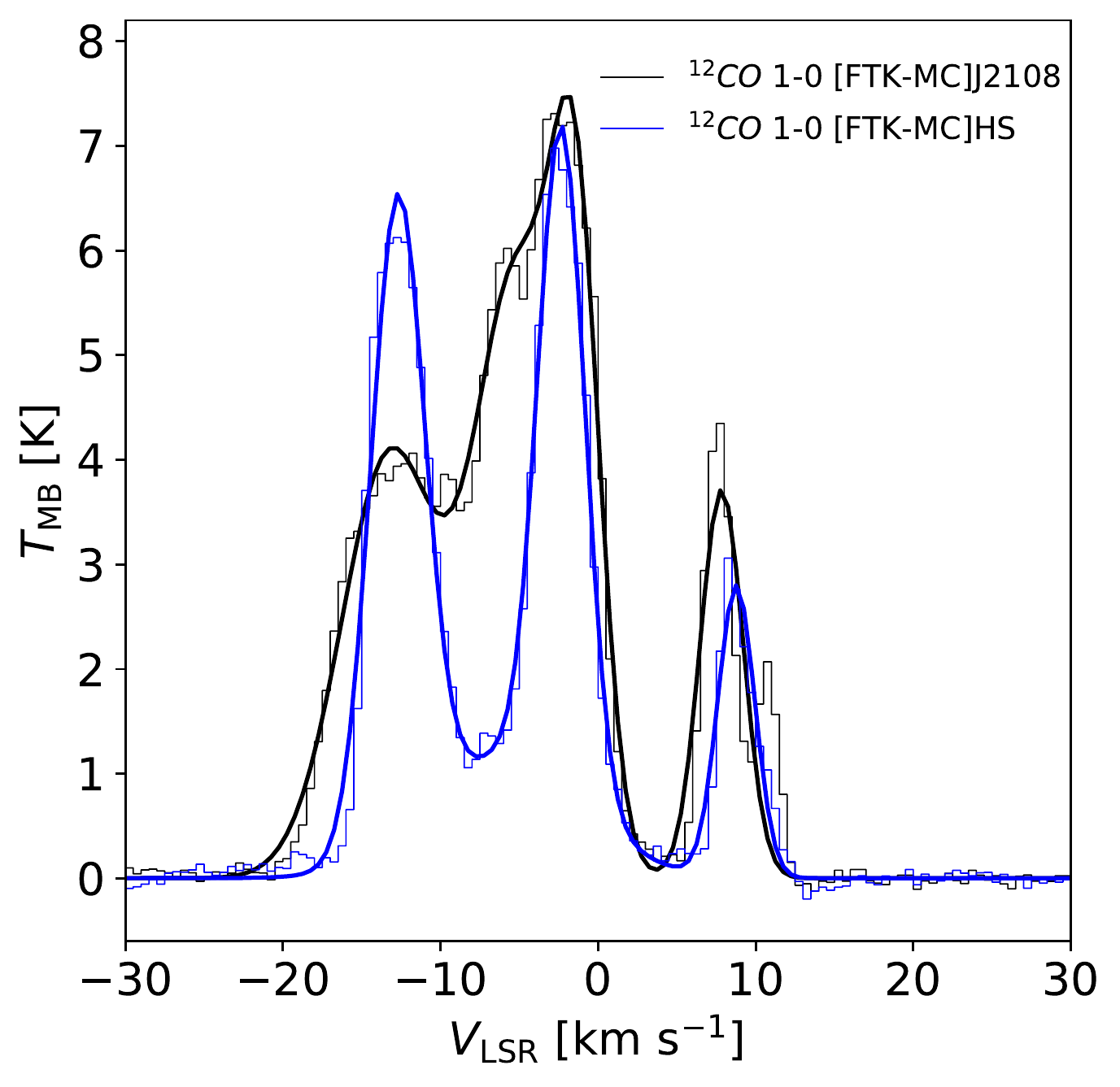}
\includegraphics[width=0.49\columnwidth]{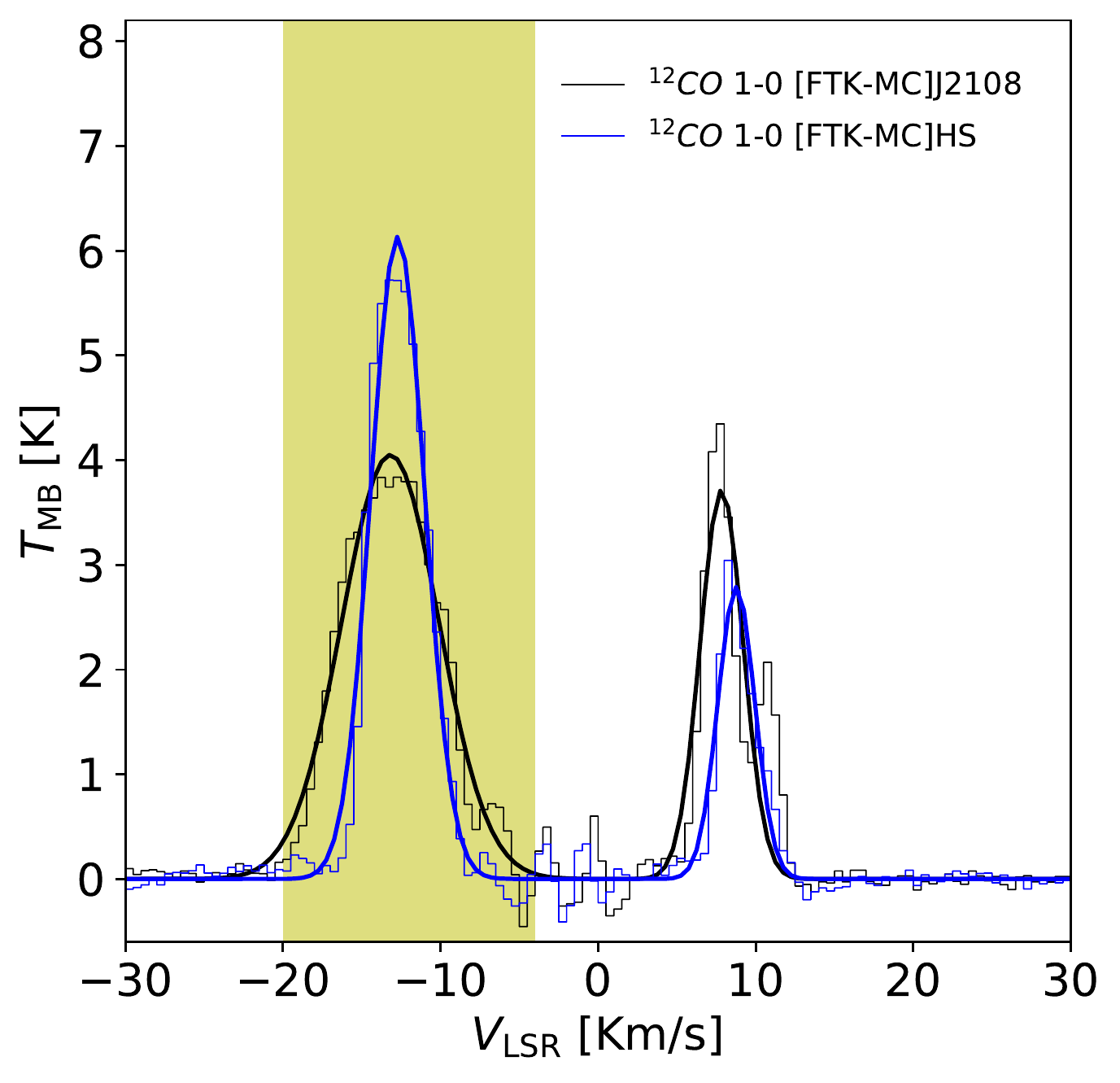}
\includegraphics[width=0.49\columnwidth]{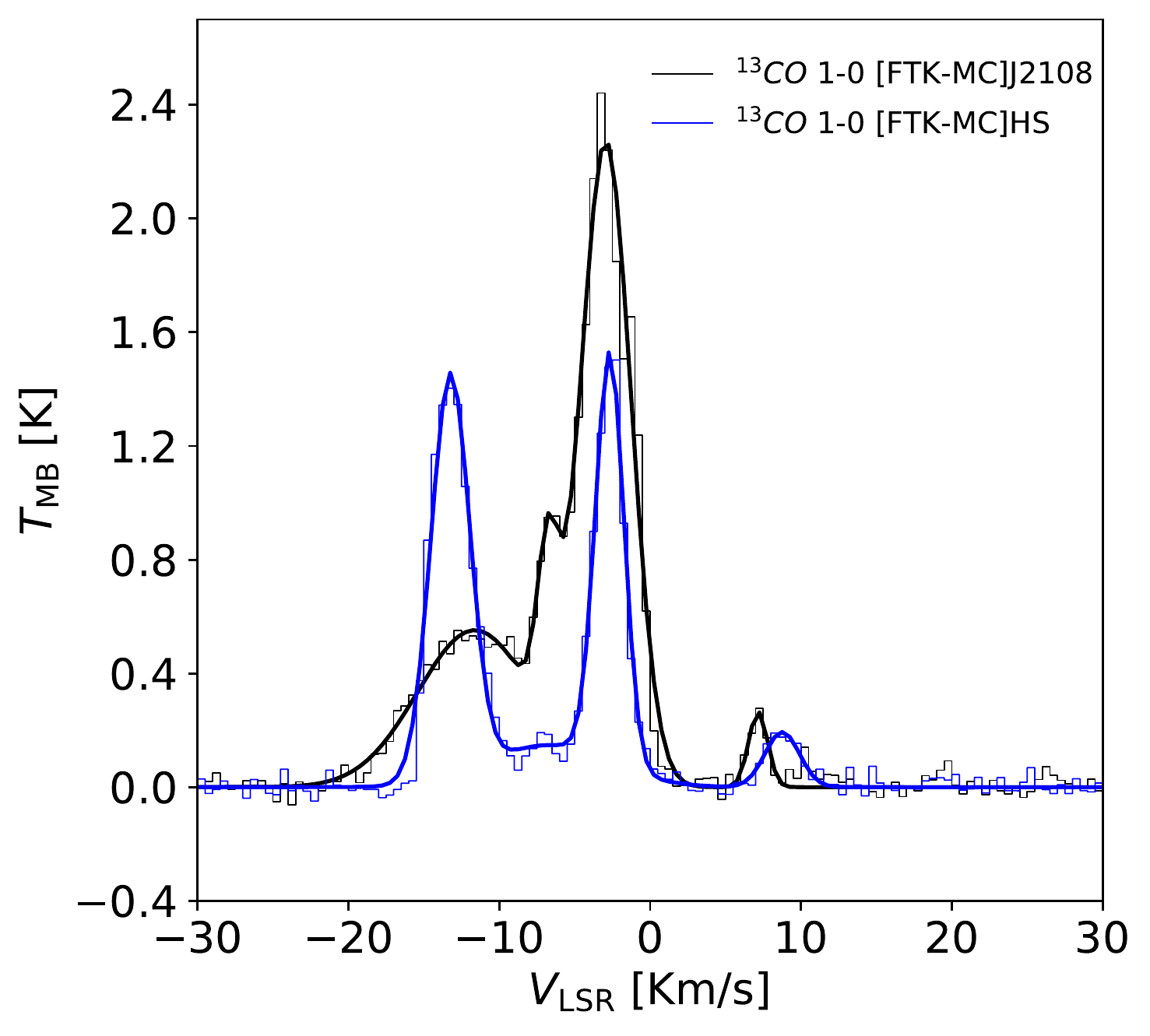}
\includegraphics[width=0.49\columnwidth]{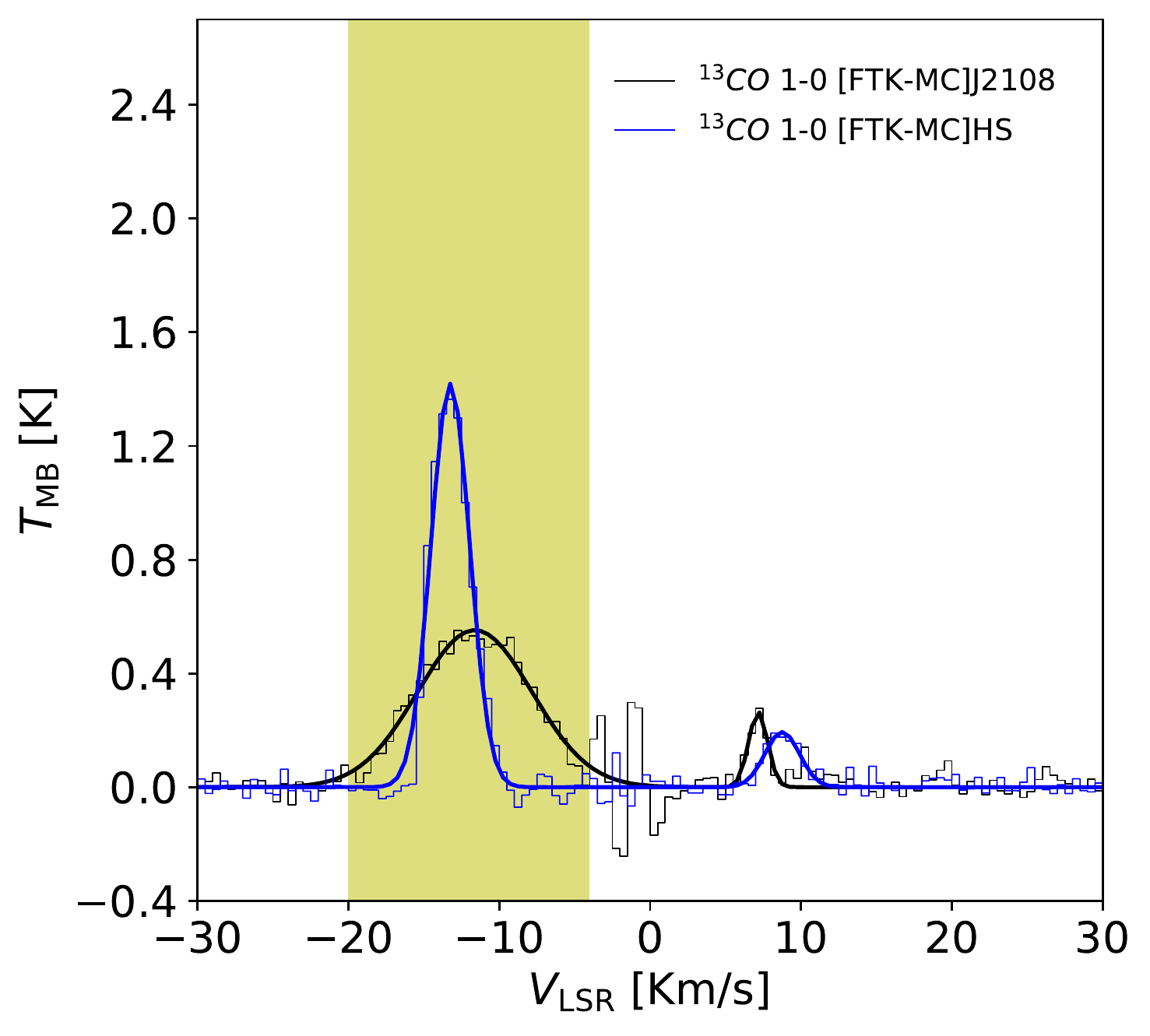}
\includegraphics[width=0.6\columnwidth]{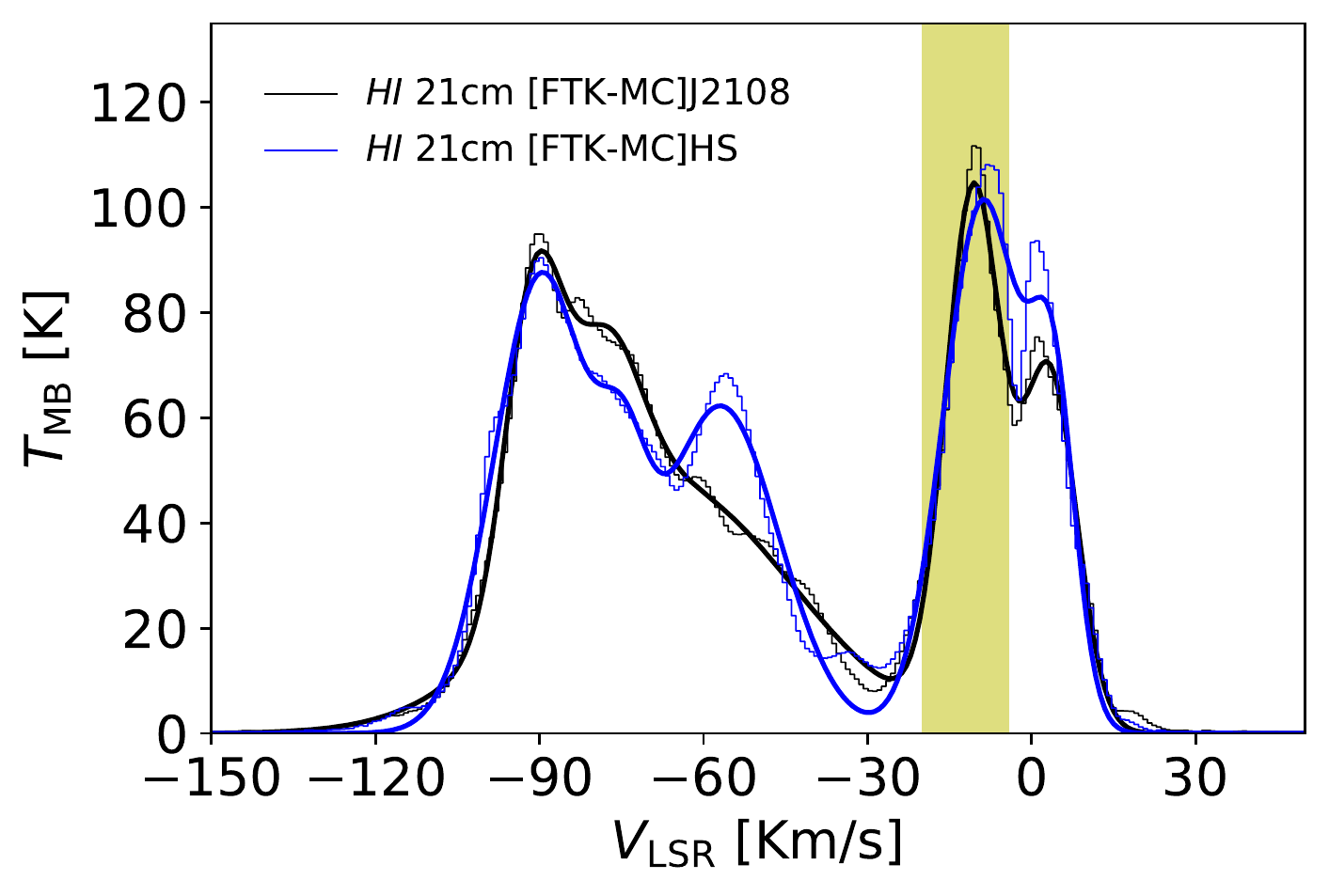}

\end{center}

\caption{Gaussian fitted extracted spectra of [FTK--MC] in $^{12}$CO (top) and $^{13}$CO (middle) for the gas associated with regions [FTK--MC]J2108 (black solid line) and [FTK--MC]HS (blue solid lines; cf. Fig.~\ref{fig:Fermi_Molecular}). The three components  C$_1$ (left), C$_2$ (middle), and C$_3$ (right) mentioned in Figs.~\ref{fig:mom0-fermi-spectrum}, \ref{fig:spectra_map}, and \ref{fig:1213CO_pv} are plotted. The corresponding spectra with the common C$_2$ (--3 km s$^{-1}$) subtracted are shown in the right panels. The DRAO HI 21 cm spectra for [FTK--MC]J2108 and [FTK--MC]HS is show at bottom. The shaded regions fill the LSR velocity range (--20 to 8 km s$^{-1}$) of [FTK--MC] for $^{12,13}$CO. They show the corresponding lines associated only with the gas of [FTK--MC].}
\label{fig:1213CO_spectra}
\end{figure}

To better determine the physical parameters of [FTK--MC] as a whole, we first analysed its two components [FTK--MC]J2108 and [FTK--MC]HS separately. The average $^{12}$CO and $^{13}$CO spectra of the [FTK--MC]J2108 and [FTK--MC]HS components are shown in Fig.~\ref{fig:1213CO_spectra} top left and middle left panels, respectively. Main beam efficiencies $\eta_{\rm MB}$ of 38.9 and 39.9\% at 115 and 110 GHz, respectively, were considered to use a main-beam brightness temperature scale. The spectra are well fitted considering four Gaussian components, two of which are related to the spectral component C$_2$. This Gaussian fit was used to extract the contribution of the spectral component C$_2$ and then to plot the corresponding reduced spectra in the top right and middle right panels for $^{12}$CO and $^{13}$CO, respectively, to better visualise and analyse component C$_1$. Next, we considered two Gaussian components to fit these reduced spectra and estimate the physical parameters of component C$_1$. In order to obtain the nucleon column density $N({\rm H}) = 2 N({\rm H_2}) + N({\rm HI})$, we show the HI 21cm spectra of [FTK--MC]J2108 and [FTK--MC]HS in the bottom panel of Fig.~\ref{fig:1213CO_spectra}. A five-component Gaussian fit was applied to the observed HI 21cm spectra, and only a V$_{\rm LSR}$ range between -20 and -4 km s$^{-1}$ (shaded range in Fig.~\ref{fig:1213CO_spectra}), which includes both [FTK--MC]J2108 and [FTK--MC]HS spectral components C$_1$, is used as the limits of the HI velocity integrated brightness temperature (see Table~\ref{table:Column2}) to determine the physical parameters.

\section{Estimation of distance to the molecular cloud [FTK--MC]}
\label{appendix:distance}

A reliable distance to the molecular cloud [FTK--MC] is essential for calculating the value of its nucleon number density, which in turn is fundamental to determine the total energy of the protons from the PeVatron that produce the observed gamma-ray emission. Using low angular-resolution observations of $^{12}$CO ($J$=1$\to$0), \cite{Cao2021b} proposed that the \object{LHAASO J2108+5157} sub-PeV emission is associated with the molecular cloud [MML2017]4607, which is located in the same direction as [FTK--MC] and has a calculated distance of 3.28 kpc \citep{Miville2017}. This distance was estimated from the rotation curve model of \citet{Brand1993}, using a position of l=92.27$^{\circ}$, b=2.77$^{\circ}$ and systemic velocity of --13.7 km s$^{-1}$. 

In order to determine the distance to [FTK--MC] we used the Bayesian distance calculator (version 2) developed by \cite{Reid2019}. To determine a probability density function of the distance this calculator uses as prior of a Bayesian analysis the probability of association to spiral arms of the Milky Way, the probability of association to near sources whose trigonometric parallaxes have been measured, and the probability of following the rotation curve of the Galaxy. We input the coordinates and systemic velocities of the molecular gas associated to the spectral components, C$_1$, C$_2$, and C$_3$ (see Fig.~\ref{fig:mom0-fermi-spectrum}), and their calculated distances are 1.62 $\pm$ 0.05 kpc (probability of 0.49), 1.61 $\pm$ 0.05 kpc (probability of 0.59), and 1.21 $\pm$ 0.26 kpc (probability of 0.47), respectively. We note that the values of the distance of components C$_1$ and C$_2$ are heavily weighted by the precise measurement of the trigonometric parallax of the source G092.69+3.08, whose distance is 1.63 kpc and is in the vicinity of the [FTK--MC] cloud. On the other hand, the value of the distance of C$_3$ reflects its possible association with the local spiral arm. If the trigonometric parallax measurement of the source G092.69+3.08 is removed from the prior of the Bayesian analysis, the distances are 2.76 $\pm$ 0.72 kpc (probability of 0.51), 1.28 $\pm$ 0.24 kpc (probability of 0.92), and 1.21 $\pm$ 0.26 kpc (probability of 0.58), for  C$_1$, C$_2$, and  C$_3$, respectively. 

As mentioned in Section \ref{sec:results_discussion}, the molecular cloud [FTK--MC] is associated to the spectral component C$_1$. From the current observations it is difficult to determine the actual distance to [FTK--MC], but it is likely that it is part of the same molecular cloud as the gas associated to the spectral component C$_2$, and both of them are located at $\sim$ 1.6 kpc, close to the source G092.69+3.08. In our analysis we only consider the C$_1$ component because of the high correlation it shows (see Fig.~\ref{fig:Fermi_Molecular}) with the gamma-ray observations of \textit{Fermi}--LAT above 2 GeV \citep{Abe2022}. We adopt the calculated distance of 1.6 $\pm$ 0.1 kpc to [FTK--MC] instead of the previous $\sim$ 3.3 kpc \citep{Cao2021b}.

\begin{figure}[!ht]
\begin{center}

\includegraphics[width=\columnwidth]{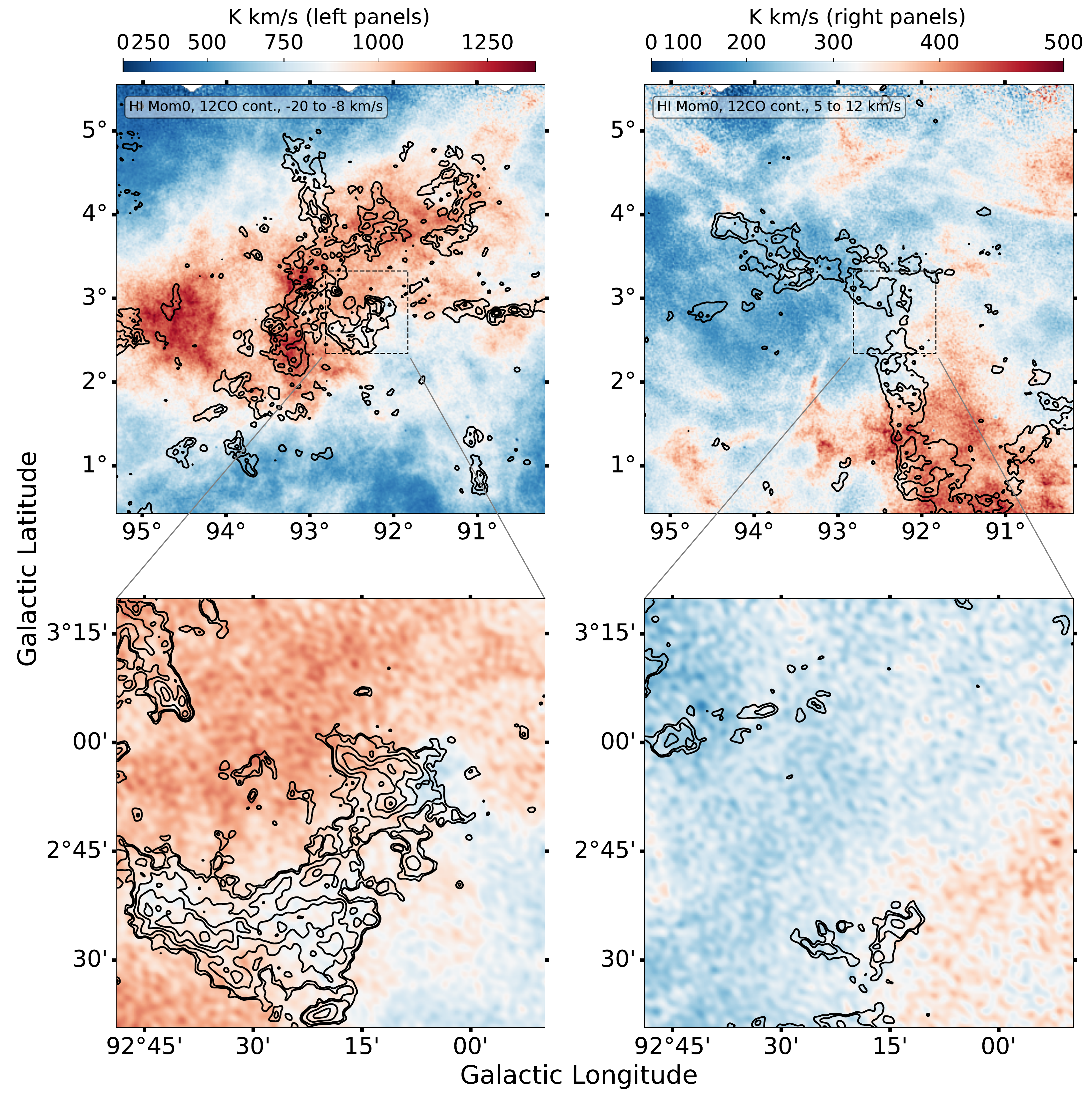}
\end{center}
\caption{OPU $^{12}$CO($J$=2$\rightarrow$1) emission in contours (Paper I) overlaid with the DRAO HI 21 cm moment-0 map (colours) in the velocity range from --20 to --8 km s$^{-1}$ (left), and 5 to 12 km s$^{-1}$ (right) at the same colour scale. The DRAO map with the Nobeyama $^{13}$CO is shown as an inset. The OPU contours are [-4, 4, 8, 12, 16, 20, 24,28,32] the rms of 1.0 K km s$^{-1}$. The Nobeyama $^{13}$CO contours are [-4,4,5,8,12,16,20] the rms of 0.5 K km s$^{-1}$.}
\label{fig:HI_13CO}
\end{figure}

Two different DRAO 21 cm moment-0 maps are shown on mesoscales in the upper panels of Figure~\ref{fig:HI_13CO}. The corresponding $^{ 12}$CO ($J$=2$\rightarrow$1) emission from OPU radio telescope are overlaid in contours \citep{delaFuente2023}. These maps consider LSR velocity ranges between --20 to --8 and 5 to 12 km s$^{-1}$, associated with C$_1$ and C$_3$ from $^{12,13}$CO emission. Two distinct HI clouds, designated H1aL and H1bR, are associated to this respective range of LSR velocities.  The corresponding $^{13}$CO  moment-0 maps are shown as insets in the bottom panels of Fig.~\ref{fig:HI_13CO}. While the molecular gas associated with C$_1$ appears to lie at the edge of H1aL, the emission associated with C$_3$ lies closer to H1bR. Therefore, we can assume that C$_3$ is not part of the molecular gas of FTK-MC, which strengthens the argument that these two spectral components are not correlated and are at different distances. Nevertheless, more sensitive HI observations with distance determinations are needed for further clarification.

\section{Tables}
\label{appendix:tables}

\begin{table*}[!ht]
	\caption{Fitted parameters of $^{12}$CO($J$=1$\rightarrow$0), $^{13}$CO($J$=1$\rightarrow$0), and HI emission via Gaussian fit for the two components of the [FTK--MC] molecular cloud (J2108 and HS gas; see Appendix \ref{appendix:distance}). The main beam (MB) averaged peak temperature ($T^p_{\rm MB}$) uncertainties are only due to rms noise. Velocity channel resolutions are used to show the V$_{\rm LSR}$ and $\Delta V$ uncertainties. For HI we refer to the shaded line in Fig.~\ref{fig:1213CO_spectra}.}
	\centering
	\begin{tabular}{cccccccc} 
		\hline
		  Component & Molecule & Spectral & Size & V$_{\rm LSR}$ & $\Delta V$ & $T^{\rm peak}_{\rm MB}$ & $\int T_{\rm MB} dv$ \\
		 Name &  & Line & [deg] & $\rm km \ s^{-1}$ & $\rm km \ s^{-1}$ & [K] & [$\rm K \ km \ s^{-1}$] \\

\hline

[FTK--MC]HS & $^{12}$CO($J$=1$\rightarrow$0) & C$_{1}$ & 0.34 $\pm$ 0.01 & --12.7 $\pm$ 0.5 & 4.0 $\pm$ 0.5 & 6.13 $\pm$ 0.66 & 26.24 $\pm$ 4.10 \\
&  & C$_{2}$ & 0.34 $\pm$ 0.01 & --2.3 $\pm$ 0.5 & 3.5 $\pm$ 0.5 & 6.30 $\pm$ 0.68 & 23.53 $\pm$ 3.21 \\
&  & C$_{3}$ & 0.34 $\pm$ 0.01 & 8.8 $\pm$ 0.5 & 2.8 $\pm$ 0.5 & 2.79 $\pm$ 0.28 & 8.21 $\pm$ 1.38 \\

& $^{13}$CO($J$=1$\rightarrow$0) & C$_{1}$ &  0.34 $\pm$ 0.01 & --13.2 $\pm$ 0.5 & 3.0 $\pm$ 0.5 & 1.42 $\pm$ 0.15 & 4.56 $\pm$ 0.63 \\
&  & C$_{2}$ &  0.34 $\pm$ 0.01 & --2.7 $\pm$ 0.5 & 2.2 $\pm$ 0.5 & 1.44 $\pm$ 0.16 & 3.42 $\pm$ 0.44 \\
&  & C$_{3}$ &  0.34 $\pm$ 0.01 & 8.8 $\pm$ 0.5 & 2.7 $\pm$ 0.5 & 0.19 $\pm$ 0.02 & 0.56 $\pm$ 0.09 \\

[FTK--MC]J2108 &  $^{12}$CO($J$=1$\rightarrow$0) & C$_{1}$ & 0.21 $\pm$ 0.01 & --13.2 $\pm$ 0.5 & 7.3 $\pm$ 0.5 & 4.05 $\pm$ 0.44 & 31.37 $\pm$ 6.10 \\
&   & C$_{2}$ & 0.21 $\pm$ 0.01 & --1.4 $\pm$ 0.5 & 3.6 $\pm$ 0.5 & 5.14 $\pm$ 1.18 & 19.74 $\pm$ 4.27 \\
&  & C$_{3}$ & 0.21 $\pm$ 0.01 & 7.8 $\pm$ 0.5 & 3.2 $\pm$ 0.5 & 3.71 $\pm$ 0.40 & 12.69 $\pm$ 1.49 \\

& $^{13}$CO($J$=1$\rightarrow$0) & C$_{1}$ &  0.21 $\pm$ 0.01 & --11.6 $\pm$ 0.5 & 8.9 $\pm$ 0.5 & 0.55 $\pm$ 0.06 & 5.25 $\pm$ 0.91 \\
&  & C$_{2}$ &  0.21 $\pm$ 0.01 & --2.9 $\pm$ 0.5 & 3.9 $\pm$ 0.5 & 2.23 $\pm$ 0.24 & 9.33 $\pm$ 0.94 \\
&  & C$_{3}$ &  0.21 $\pm$ 0.01 & 7.2 $\pm$ 0.5 & 1.5 $\pm$ 0.5 & 0.27 $\pm$ 0.03 & 0.42 $\pm$ 0.08 \\

\hline
	\end{tabular}
	\label{table:fit_parameters}
\end{table*}

\begin{table*}
	\caption{\centering Observational parameters of the components of the [FTK--MC] molecular cloud}
	\centering
	\begin{tabular}{ccccccc} 
		\hline
		Component & Molecular & Diameter & $T_{\rm MB}^{\rm peak}$ & $\int T_{\rm MB} dV$ & $T_{\rm ex}$ & $\tau$$^{a}$  \\
		 Name & Line & [deg] & [K] &[K km s$^{-1}$] & [K] &    \\
		\hline
[FTK--MC]HS & $^{12}$CO($J$=1$\rightarrow$0) & 0.34 $\pm$ 0.01 & 6.13 $\pm$ 0.66 & 26.24 $\pm$ 4.10 & 9.47 $\pm$ 0.68 & 15.64 $\pm$ 2.72 \\
 & $^{13}$CO($J$=1$\rightarrow$0) & 0.34 $\pm$ 0.01 & 1.42 $\pm$ 0.15 & 4.56 $\pm$ 0.63 & 9.47 $\pm$ 0.68 & 0.26 $\pm$ 0.05 \\
 & HI (21 cm) & 0.34 $\pm$ 0.01 & -- & 1243.61 $\pm$ 33.78 & -- & -- \\
 
[FTK--MC]J2108 & $^{12}$CO($J$=1$\rightarrow$0) & 0.21 $\pm$ 0.01 & 4.05 $\pm$ 0.44 & 31.37 $\pm$ 6.10 & 7.31 $\pm$ 0.46 & 8.70 $\pm$ 1.43 \\
 & $^{13}$CO($J$=1$\rightarrow$0) & 0.21 $\pm$ 0.01 & 0.55 $\pm$ 0.06 & 5.25 $\pm$ 0.91 & 7.31 $\pm$ 0.46 & 0.14 $\pm$ 0.02 \\
 & HI (21 cm) & 0.21 $\pm$ 0.01 & -- & 1207.18 $\pm$ 19.50 & -- & -- \\
\hline
	\end{tabular}\\
{\small \raggedright $^{a}$ Optical depth of the $^{12}$CO and $^{13}$CO and molecular line emission.   
        \par}       
	\label{table:Column1}

\end{table*}

\begin{table*}
	\centering
 \caption{Physical parameters of the [FTK--MC] molecular cloud $^{a}$}
 \centering
  	\begin{tabular}{cccccccc} 
		\hline
		$N(\rm ^{13}CO)$ & $N(\rm HI)$ &  $N(\rm H_2)$$^{b}$ & $n({\rm HI)}$ & $n({\rm H_2})$& $M_{\rm vir}({\rm H_2})$$^{b}$ & $M({\rm H_2})$  & $M(\rm HI + H_2)$\\
  
  [10$^{15}$ cm$^{-2}$] & [$10^{21}$ cm$^{-2}$] & [$10^{21}$cm$^{-2}$] & [$\rm cm^{-3}$]& [$\rm cm^{-3}$] & [$10^4 M_\odot$] & [$10^3 M_\odot$] & [$10^3 M_\odot$]\\
		\hline
	    4.0 $\pm$ 0.4 & 2.2 $\pm$ 0.3 & 2.0 $\pm$ 1.0 &  48 $\pm$ 7 &  43 $\pm$ 22 & 5.8 $\pm$ 1.0 & 5.3 $\pm$ 2.8 & 7.5 $\pm$ 2.9\\
		\hline
	\end{tabular}
 
        {\small \raggedright $^{a}$ For the calculations of the physical parameters a distance of 1.6 $\pm$ 0.1 kpc was assumed. The angular size of the [FTK--MC] molecular cloud is considered to be 0.55 $\pm$ 0.02 degrees, which is the sum of the sizes of the two individual components [FTK--MC]HS and [FTK--MC]J2108.\\
        $^{b}$For the virial mass calculation an average of the line-widths of C$_{1}$ of the $^{13}$CO($J$=1$\rightarrow$0) emission from [FTK--MC]HS and [FTK--MC]J2108 was used.  
        \par}
	\label{table:Column2}
\end{table*}

\begin{table*}
   \centering
    \caption{Parameters and results of the hadronic model of Naima for the [FTK--MC] molecular cloud}
    \begin{tabular}{cccccc} \hline
        Distance &$N(\rm H)$$^{a}$& n(H)$^{a}$ & Size & $W_p$ & Cutoff  \\ 
    
        [kpc] &  [$10^{21}$ cm$^{-2}$]&[cm$^{-3}$] & [degree] & [$10^{47}$erg] & [TeV]   \\ \hline
        1.6$\pm$0.1 &6.2$\pm$2.1 & 133$\pm$45 & 0.55$\pm$0.02 & $1.1^{+0.6}_{-0.4}$ & $600^{+400}_{-200}$  \\
      \hline
    \end{tabular}
    
    {\small \raggedright $^{a}$ The column and number density of nucleons is calculated as N(H) = 2N(H$_2$) + N(HI) and n(H) = 2n(H$_2$) + n(HI), respectively.
        \par}
    \label{tab:hadron}     
\end{table*}

\end{appendix}

\end{document}